\documentclass{aa}
\usepackage{graphics}

\begin{document}

%   \thesaurus{10         % A&A Section10: The Galaxy
%	      (09.03.1; %ISM: clouds,
%	       09.05.1; %ISM: evolution,
%	       09.08.1; %ISM: HII regions
%	       10.03.1) %Galaxy: center
%             }
%
   \title{A cloudy model for the Circumnuclear Disk in the Galactic Centre}

   \titlerunning{A cloudy model for the CND}

   \author{B.~Vollmer\inst{1,2,3} \and  W.J.~Duschl\inst{1,2}}

   \offprints{B.~Vollmer, e-mail: bvollmer@mpifr-bonn.mpg.de}

   \institute{Institut f\"ur Theoretische
              Astrophysik der Universit\"at Heidelberg, Tiergartenstra{\ss}e 15,
              D-69121 Heidelberg, Germany. \and
              Max-Planck-Institut f\"ur Radioastronomie, Auf dem H\"ugel 69,
   	      D-53121 Bonn, Germany. \and
	      Observatoire de Meudon, DAEC,
              UMR 8631, CNRS et Universit\'e Paris 7,
	      F-92195 Meudon Cedex, France.}

   \date{Received / Accepted}

\abstract{
We present a first attempt to construct an analytic model for a clumped
gas and dust disk and apply it to the Galactic Centre. The clumps
are described as isothermal spheres partially ionized by the external
UV radiation field. The disk structure formed by the clouds is described
as a quasi standard continuous accretion disk using adequately averaged
parameters of the discrete cloud model. 
The viscosity in the Circumnuclear Disk is due to partially inelastic 
cloud--cloud collisions. We find two different solutions for the set of equations 
corresponding to two stable cloud regimes: (i) the observed molecular clouds and 
(ii) much lighter and smaller clouds which correspond to the stripped cores of
the observed clouds.
It is shown that the resulting physical characteristics of the heavy clouds
and the disk are in very good agreement with all comparable observations 
at multiple wavelengths. A mass accretion rate of 
$\dot{M}\simeq 10^{-4}$\,M$_{\odot}$\,yr$^{-1}$ for the isolated Circumnuclear Disk
is inferred. We propose that the Circumnuclear Disk has a much 
longer lifetime ($\sim 10^{7}$ yr) than previously assumed.
}

\maketitle

\section{Introduction}

The Galactic Centre is surrounded by a large number of gas and dust clouds
forming a thick disk ({\it Circumnuclear Disk CND}) up
to a radius of $\sim$7 pc\footnote{We assume 8.5 kpc for the distance to 
the Galactic Centre.}. This disk was discovered by 
Becklin, Gatley, \& Werner (1982) interpreting their FIR data as a tilted 
dust ring.
It has its minimum emission towards the central compact radio source
SgrA$^{*}$. Subsequently, the CND was investigated by several authors
observing the emission of dust, molecules and atoms.
\begin{itemize}
\item
Dust: Mezger et al. (1989), Davidson et al. (1992), Dent et al. (1993), 
Telesco et al. (1996)
\item
Molecules: Gatley et al. (1986) (H$_{2}$), Serabyn et al. (1986) (CO, CS), 
G\"{u}sten et al. (1987) (HCN),
DePoy et al. (1989) (H$_{2}$),  Sutton et al. (1990) (CO),
 Jackson et al. (1993) (HCN), Marr, Wright, \& Backer (1993) (HCN)
\item 
Atoms: Lutgen et al. (1986) (C{\sc ii}), Jackson et al. (1993) (O{\sc i}) 
\end{itemize}      
They concluded that the CND has a hydrogen mass of a few 10$^{4}$ M$_{\odot}$.
The disk is very clumpy with an estimated area filling factor of 
$\Phi_{\rm A} \sim 0.1$ and a volume filling factor of 
$\Phi_{\rm V} \sim 0.01$. 
The clumps have densities of several 10$^{5}$ cm$^{-3}$,
radii of $\sim$0.1 pc and gas temperatures $\geq$100 K. A typical dust clump 
has A$_{\rm V} > 30^{\rm m}$ and M$_{\rm H} \sim$30 M$_{\odot}$.
The physical parameters of the clouds in the central 2 pc are listed in 
Jackson et al. (1993). 
They are partially ionized by the radiation of the
central He{\sc i} star cluster (see e.g. Genzel et al. 1996).  
A low density ionized gas
with a density of $\sim$10$^{3}$ cm$^{-3}$ (Erickson et al. 1994) and
a temperature of a typical H{\sc ii} region of T$\sim$7000-8000 K
surrounds the clouds.
The vertical thickness of the CND increases from $\sim$0.5 pc at
a radius of 2 pc to $\sim$2 pc at 7 pc.
The disk rotates with a velocity of $\sim$100 km\,s$^{-1}$ which
corresponds to a Keplerian velocity around a central object of
several 10$^{6}$ M$_{\odot}$ and has a velocity dispersion of $\sim$ 
30 km\,s$^{-1}$.  
It is inclined by an angle $i \sim 20^{\rm o} - 30^{\rm o}$ relative to
the line of sight (LOS). 
Due to its clumped structure, the CND is usually assumed to be a rather
short-lived transient feature ($\sim 10^{5}$ yr) (G\"usten et al. 1987).

At a distance of $\sim$1.7 pc from Sgr A$^{*}$ it has a sharply defined 
inner edge (Marr et al. 1993).
The dust and molecular line emission drops there by an order of magnitude.
This defines the outer limit of the {\it Central Cavity (CC)} which contains 
the H{\sc ii} region Sgr A West. It was investigated in detail by
Lo \& Claussen, Lacy et al. (1991), Roberts \& Goss (1993), Lacy (1994).

Three dimensional kinematical models were made by  
Davidson et al. (1992) for the
dust emission, Marshall \& Lasenby (1994) for the molecular line
emission, and Vollmer \& Duschl (2000) for the ionized component. 
They conclude that there is a single plane in which the major parts of the 
CND is located.

As yet, there are only few attempts to construct a physical model for
the CND. Wardle \& K\"{o}nigl (1990, 1993) investigated a continuous 
smooth disk model
for the CND including the magnetic field. They succeeded in explaining the 
dust polarization observed by Hildebrand et al. (1993).  
On the other hand, Krolik  \& Begelman (1988) constructed a clumpy disk model
for AGNs where cloud-cloud collisions are responsible for the energy and 
momentum transport and thus for the viscosity.
Cloud--cloud collisions dissipate orbital energy resulting a net
inward drift of the clouds.
If the angular momentum of one cloud is low enough it can be `captured by the
central object, providing its accretion fuel' (Krolik  \& Begelman 1988). 
They mentioned that the clouds observed by Genzel et al. (1985)
have a column density comparable to the Jeans column density.
However, they conclude that selfgravity cannot overcome tidal shear.
 
Shlosman \& Begelman (1987) pointed out that a necessary condition
for the fragmentation of an externally heated disk is that the
cooling time must be shorter than the orbital time. They argued that
this is often the case for disks in which the temperature
is regulated by dust. Furthermore, Shlosman et al. (1990) discussed
the possibility of fueling an AGN by the means of a cloudy disk.

Given the area and volume filling factor of the CND, it is 
clear that a continuous, smooth molecular disk is ruled out.
A model of a clumpy disk is needed. In this paper we present
such a model. First we describe the outlines
of the model disk physics (section 2). We then give a description 
of the viscosity in the disk and discuss the energy dissipation mechanism
in section 3. We present the model for a partially
ionized globule (PIG), give the equations for the disk, and 
treat the physical conditions in section 4. The results and the
verification of the assumptions are shown in section 5. Section 6
treats the influence of the tidal shear on the clouds. 
We discuss the mass accretion rate in section 7.
The conclusions are given in section 8.

\section{The outlines of the model}

We assume that during a short accretion event ($\Delta t \sim 10^{6}$ yr)
an amount of gas of several 10$^{4}$ M$_{\odot}$ is driven into the
Galactic Centre region at distances less than 10 pc. 
The accreted mass is not uniform but
has density fluctuations. This clumpy medium is exposed to the ambient
UV radiation field due to the population of young O/B stars in the Galactic
Centre. Low density regions of the infalling gas are evaporated 
rapidly while regions of higher density stay molecular and are heated 
to an equilibrium temperature during less than an orbital period. 
In this way gas clouds
of different masses and different sizes are formed. At a given distance from
the Galactic Centre only clouds with a central density high enough
to resist tidal shear can survive. In these clouds the thermal pressure
is balanced by gravitation. They are assumed to be isothermal.
Since the UV radiation comes mainly from the Galactic Centre
their radius in this direction is given by the location 
of the ionization front resulting from the incident radiation.
The radius of the opposite side is given by the pressure of the
ionized gas which fills the space between the clouds.
During the infall the clouds have frequent partially inelastic collisions.
These collisions are highly dissipative 
because the cooling time is much shorter
than the duration of the collision (Krolik \& Begelman 1988). 
These collisions can lead to cloud fragmentation, mass exchange, or
coalescence depending on the mass ratio and velocity difference
of the colliding clouds. 
If the initial infalling gas has a total angular moment which is 
not zero these collisions will lead to the formation of a disk structure, 
corresponding to a clumpy accretion disk, i.e. the CND. 
Within this disk structure clouds lose
orbital energy during dissipative collisions. This results in a net inward 
drift of the clouds, i.e. angular momentum is transported due to cloud--cloud
collisions (Krolik \& Begelman 1988, Ozernoy, Fridman, \& Biermann
1998).  

We adopt two different views to model the CND as a clumpy accretion disk
as described above. First, the small-scale aspect where the gas clouds
are treated as isothermal selfgravitating spheres.
For each distance to the Galactic Centre the cloud structure is 
calculated with a given central density, temperature, and external pressure
due to the ionization front. 
Second, the large-scale aspect where the cloud
distribution is smoothed out over the whole disk resulting in a 
continuous accretion disk model. Both models share the same temperature 
distribution and the same UV radiation field with respect to their
distance to the Galactic Centre. The connection between the small-scale
and the large-scale model lies in the link between the central density of the
clouds and that of the accretion disk. In addition, the disk viscosity
due to partially inelastic collisions is taken into account in the
large-scale model.  

\subsection{The gas clouds} 
 
\begin{figure}
	\resizebox{\hsize}{!}{\includegraphics{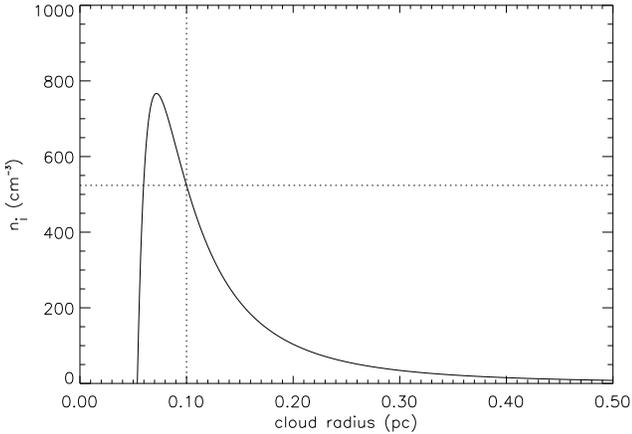}}
      \caption{ \label{fig:globule} 
	The electron density corresponding to the boundary pressure
	of the cloud as a function of the cloud radius.
	The dotted lines indicate the electron density at a
	typical cloud radius of 0.1\,pc.
}
\end{figure}

The gas clouds are
exposed to the radiation field of the central He{\sc i} star 
cluster, which ionizes the illuminated surface. The gas
in the ionization front
flows away from the clump and fills up the whole volume
of the disk providing an outer pressure at the cloud surface
which is not illuminated. The external radiation field is also
responsible for the gas temperature within the cloud, which
is assumed to be constant. Consequently, the cloud is modeled
as an isothermal sphere whose boundary is given at the illuminated
side by the location of the ionization front and at the
shadowed side by the pressure of the ionized intercloud gas.
For a quantitative estimate we can
solve the virial theorem for the surface pressure $P$:
\begin{equation}
P=\frac{c_{v}M_{\rm cl} T}{2\pi r_{\rm cl}^{3}}-
\frac{\Theta G M_{\rm cl}^{2}}{4 \pi r_{\rm cl}^{4}}\ ,
\end{equation} \label{eq:a1}
where $c_{v}$ is the specific heat, $M_{\rm cl}$ the cloud mass,
$T$ the temperature inside the cloud, $r_{\rm cl}$ its radius, and $\Theta$
is a factor of order one. 
The outer pressure can be translated into an electron density,
taking $T$=7000\,K for the intercloud medium (Roberts \& Goss 1993).
We set $M_{\rm cl}=15$\,M$_{\odot}$ and $T$=100\,K. The deduced electron density is
plotted in Fig.~\ref{fig:globule}.
The electron density at a radius of 0.1\,pc
is of the order of 500\,cm$^{-3}$ which compares very well with
the observed value (e.g. Erickson et al. 1994) considering the approximate
character of the model. 

The CND consists of $\sim$1000 gas clouds. Each of these clouds has a mass
of $\sim$30 M$_{\odot}$ (see e.g. Jackson et al. 1993).

\subsection{The accretion disk}

The clouds can have partially inelastic collisions, which are responsible
for the energy dissipation and the transport of angular momentum,
i.e. the viscosity (Ozernoy et al. 1998). 
The standard viscosity description does not apply in this case,
because the turbulence in the CND is supersonic. 
Therefore, we derive a new viscosity description, which is based on Kolmogorov's theory.
This viscosity depends on the disk height and the energy dissipation rate.
At the smallest scales the energy is radiated away by infrared line emission. 
Thus, the disk is described following
the standard equations for a smooth continuous disk (see e.g. Pringle 1981)
with the modified viscosity description. In this picture the different
components (neutral and ionized) of the discrete cloud model are completely 
mixed and the disk central density is directly related to the
clouds' central density at a given distance from the Galactic Centre.
   
\section{The turbulent viscosity}

In a turbulent medium kinetic energy is transferred from
large scale structures to small scale structure practically without
losing energy. So there is a constant energy flux from large scales
to small scales where the energy is finally dissipated. 
Since the velocity dispersion ($\Delta v \sim 20$ km\,s$^{-1}$) 
within the disk is more important than the
shear, there is no preferred transfer direction. The turbulence can
thus be assumed as isotropic. In this case the similarity theory
of Kolmogorov applies (see e.g. Landau \& Lifschitz 1959).
The assumption of a universal Kolmogorov equilibrium implies
that the kinetic energy spectrum of the turbulence depends only on 
the energy dissipation rate per mass unit $\epsilon$ and the characteristic
size of the turbulent eddy $l \simeq \frac{1}{k}$, where $k$ is the 
wave number.
The kinetic energy $E(k)$ is related to the mean kinetic energy in
the following way:
\begin{equation}
\frac{1}{2}\langle {\bf u}(x)^{2} \rangle = \int_{0}^{+\infty} E(k) dk\ ,
\end{equation} \label{eq:a2}
where ${\bf u}$ is the velocity of the medium. Kolmogorov's theory yields 
\begin{equation}
E(k)=C \epsilon^{\frac{2}{3}} k^{-\frac{5}{3}}\ ,
\end{equation} \label{eq:spectrum}
where C is a constant of the order of unity.
Considering a schematic energy spectrum given by
$E(k)=0$, for $k < k_{\rm turb}$ and for $k > k_{d}$ and by equation 
(\ref{eq:spectrum}) for $k_{\rm turb} < k < k_{d}$, 
we can derive expressions for
the dissipative scale length $l_{d} \simeq k_{d}^{-1}$, the large scale
turbulence scale $l_{\rm turb} \simeq k_{\rm turb}^{-1}$, and the large scale
velocity $v_{\rm turb}$:
\begin{equation}
l_{d}=(\nu^{3}/\epsilon)^{\frac{1}{4}}\ ,
\end{equation} \label{eq:a3}
where $\nu$ is the large scale viscosity due to turbulence;
\begin{equation}
v_{\rm turb}^{2}=\langle {\bf u}^{2} \rangle \simeq \epsilon^{\frac{2}{3}} k_{\rm turb}^
{-\frac{2}{3}}\ .
\end{equation} \label{eq:a4}
This leads to a relation between the two length scales and the turbulent
Reynolds number $Re = v_{\rm turb}\cdot l_{\rm turb}/\nu$
\begin{equation}
l_{\rm turb} \simeq Re^{\frac{3}{4}} l_{d}\ .
\end{equation} \label{eq:a5}
Consequently the turbulent large scale viscosity $\nu$ is given by
\begin{equation}
\nu \simeq \frac{1}{Re}\epsilon^{\frac{1}{3}} l_{\rm turb}^{\frac{4}{3}}\ .
\end{equation} \label{eq:a6}
The macroscopic pressure due to the turbulence is given by
\begin{equation}
p \simeq \rho \cdot v_{\rm turb}^{2} \simeq \rho \cdot (\epsilon \cdot l_{\rm turb})
^{\frac{2}{3}}\ ,
\end{equation} \label{eq:a7}
where $\rho$ is the overall density in the turbulent medium. 
We recall that, since there is only negligible energy loss
during the energy transfer from large scales to small scales, 
the energy dissipation rate per unit mass for large structures
is the same as that for small structures and is determined
by the actual physical dissipation process in the smallest structures 
(heating and radiation). If we can identify this dissipative
process in a given observed system, it should be possible
to draw conclusions about the large-scale turbulence.
In the next section we give a possible radiative dissipation mechanism
in order to get an analytic expression for $\epsilon$.

\subsection{Energy dissipation by radiative cooling}

The clouds in the CND are subject to partially inelastic collisions where they
lose kinetic energy. During the collision the clouds' gas is heated, 
radiating away its excess energy in the infrared lines of O{\sc i}, 
C{\sc ii}, and excited H$_{2}$ emerging from the hot outer layer of the cloud. 
During a collision a shock front is formed heating the gas in
the interacting region to $\sim$1000 K. In a detailed model one has
to account for this enhanced temperature within a small region.  
In order to treat the energy dissipation in our large-scale model
we are not interested in the details of these non-stationary hot layers
but in a smooth stationary dissipation rate over the whole disk. 
We therefore make the assumption that the time and space averaged dissipation 
rate in these hot layers can be approximated by the continuous infrared 
line emission of the smoothed large scale disk. 
This point will be further discussed in section 5.3.

At temperatures of $\sim$200 K and densities of several $10^{4}$ cm$^{-3}$,
which represent the averaged values of our disk model,
the O{\sc i} and C{\sc ii} line intensities are comparable 
(see e.g. Tielens \& Hollenbach 1985). Here, we will only take into account 
the C{\sc ii} line at 158 $\mu$m, assuming that
it is representative for the radiative dissipation of the turbulent
energy in the disk. 

The cooling function of the C{\sc ii} line at 158 $\mu$m is (Spitzer 1978) 
\begin{equation}
\Lambda_{\rm CII} = \frac{\Delta E}{\Delta V_{\rm PDR}\,\Delta t}=7.9\,10^{-27}d_{\rm C}e^{-92/T}n_{\rm H}^{2} 
\ {\rm erg}\, {\rm cm}^{-3}\, {\rm s}^{-1}\ ,
\end{equation} \label{eq:a8}
where $\Delta V_{\rm PDR}$ is the volume of the photodissociation region (PDR), 
$T$ is the disk temperature, $n_{\rm H}$ the hydrogen density in the disk 
and $d_{\rm C}=\frac{n_{\rm C}/n_{\rm H}}{(n_{\rm C}/n_{\rm H})_{\odot}}=1$ 
is the fraction of atomic carbon abundance in the disk
with respect to the solar one 
($(n_{\rm C}/n_{\rm H})_{\odot} \simeq 3\,10^{-4}$).
This is valid because at the given density ($n > n_{\rm crit}
\simeq 3\, 10^{3}$ cm$^{-3}$) and at a temperature beyond 200 K 
the level populations are simply determined by their statistical 
weights (i.e. LTE) (Wolfire, Tielens, \& Hollenbach 1990).

The radiative cooling takes place in the outer layers of the clouds, i.e. in the PDR,
which represent only a small fraction of the whole cloud volume.
In the large-scale model the ionized, atomic, and molecular phases are mixed.
If the total cloud volume is $\Delta V$, the modified cooling function is 
\begin{equation}
\Lambda_{\rm CII}^{\rm mod}= \frac{\Delta E}{\Delta V\,\Delta t} =
7.9\,10^{-27} \eta e^{-92/T}n_{\rm H}^{2} \ {\rm erg}\, {\rm cm}^{-3}\, {\rm s}^{-1}\ ,
\end{equation}
where $\eta = \Delta V_{\rm PDR}/\Delta V \simeq M_{\rm atomic}/M_{\rm tot} 
\simeq N_{\rm atomic}/{\overline N_{\rm tot}}$.
$M_{\rm atomic}$ is the mass of atomic gas located in the PDR, $M_{\rm tot}$ is its total mass,
$N_{\rm atomic}$ is the column density of atomic gas, and
$\overline{N_{\rm tot}}=m_{\rm p}^{-1}\rho\,d$ is the total column 
density of a cloud of diameter $d$.

The column density of the PDR is determined by H$_{2}$/CO self-shielding 
or dust absorption. As the C$^{+}$/C/CO transition region
is difficult to determine, we decided to use only a normalized value of the 
H/H$_{2}$ transition region.  
For the self-shielding its value $\eta^{\rm H_{2}}$ follows 
from the balance of formation and destruction of 
H$_{2}$ (Burton, Hollenbach, \& Tielens 1990). 
\begin{equation}
N_{\rm atomic}\simeq \frac{2 I_{0}^{2} G_{0}^{2} \beta^{2}}{\langle R \rangle
^{2}n_{\rm surface}^{2}}\ ,
\end{equation} \label{eq:a9}
where $I_{0} = 6\, 10^{-11}$ s$^{-1}$ is the unshielded 
photodissociation rate, $\beta = 4\,10^{5}$ cm$^{-1}$ is the 
selfshielding parameter, $\langle R \rangle = 3\,10^{-17}$ cm$^{3}$\,s$^{-1}$ 
is the H$_{2}$ formation rate coefficient, $n_{\rm surface}$ is
the hydrogen density at the cloud surface, and $G_{0}$ is the FUV radiation 
field in units of the equivalent Habing (1968) flux of 
$1.6\,10^{-3}$ erg\,cm$^{-2}$\,s$^{-1}$.
This gives $\eta^{\rm H_{2}}=\gamma\ (N_{\rm atomic}/
{\overline N_{\rm tot}})$. The factor $\gamma$ is determined by the
condition that the self-shielding becomes the dominant process for
$n_{\rm surface}/G_{0} > 100$ cm$^{-3}$ (Burton et al. 1990).
For the dust absorption the column density is fixed by the UV optical
depth $\tau_{\rm UV}\sim 1$. This corresponds to a constant column density
of $N_{\rm atomic}^{\rm dust}\simeq 2\, 10^{21}$ cm$^{-2}$, resulting in 
$\eta^{\rm dust}=N_{\rm atomic}^{\rm dust}/{\overline N_{\rm tot}}$.
The total volume fraction of atomic gas is obtained by 
$\frac{1}{\eta^{\rm tot}}=\frac{1}{\eta^{\rm H_{2}}}+
\frac{1}{\eta^{\rm dust}}$.

This leads to an energy dissipation rate per mass unit of
\begin{equation}
\epsilon = \Lambda_{\rm CII}^{\rm mod} / \rho = 
7.9\,10^{-27}m_{\rm p}^{-2}\eta^{\rm tot}e^{-92/T}\rho=:\xi\cdot \rho\ ,
\end{equation}
which can be used in the expression for the viscosity and
the turbulent pressure derived above.

\section{The detailed model}

\subsection{Partially ionized globules (PIG)}

\begin{figure}
	\resizebox{\hsize}{!}{\includegraphics{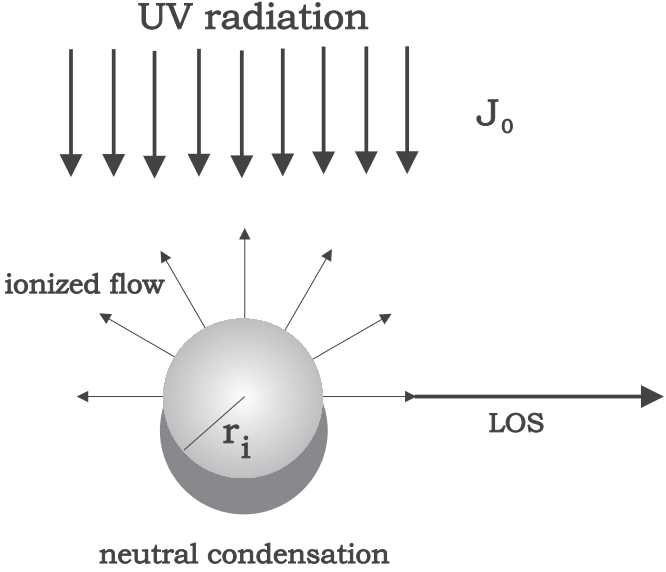}}
      \caption{Illustration of the in the text described PIG model.
	The UV radiation comes from the direction of the Galactic Centre.
	The lower boundary is determined by the gas pressure of the 
	surrounding ionized low density gas.}
	\label{fig:globulenew}
\end{figure}

The clouds embedded in the H{\sc ii} region Sgr A West 
and illuminated by the ambient radiation field are assumed to be
 spherically symmetric, isothermal, and selfgravitating.
We follow the model of Dyson (1968). The incident UV radiation 
ionizes the outer rim and produces an ionization front. As
only photons above 13.6 eV ionize the hydrogen atoms in
the ionization front, the FUV field can penetrate further into
the cloud. There, these photons cause the photodissociation of the H$_{2}$
molecules. Both the ionization front and the dissociation front
are governed by dust absorption and self-shielding effects.
In addition, the PDR is preceded by a shock front.
Thus, the cloud structure has
three components, the inner molecular core, the predominantly
atomic PDR and the outer ionization front.
Here we are only interested in the ionization front, which is assumed to be
quasi stationary and D-critical where the velocity of the shock front is
approximately the neutral gas sound velocity squared divided
by twice the ionized gas sound velocity (see e.g. Spitzer 1978).
Due to the much higher temperature
in the ionization front, the electrons are torn away from the cloud,
giving rise to an effective mass loss.
Across the front, the following jump conditions apply:
\begin{equation}
v_{n}=\frac{c_{n}^{2}}{2c_{i}}; \ \ \ \ v_{i}=c_{i};\ \ \ \ \frac{\rho_{i}}
{\rho_{n}}=\frac{c_{n}^{2}}{2c_{i}^{2}}\ ,
\end{equation} \label{eq:a12}
where $v$ is the gas velocity and $c$ is the isothermal sound speed.
The index $n$ stands for the neutral, the index $i$ for the ionized gas 
component which is located directly in front of the ionization front.      
The outward electron flux is assumed to be isothermal and stationary
because the ionization front moves very slowly into the cloud.
The justification for all these assumptions is given in section 5.3.
The solution for a spherically symmetric, stationary flux is given by
\begin{equation}
\frac{r}{r_{i}}=(\frac{c_{i}}{v})^{\frac{1}{2}}\exp(\frac{c_{i}^{2}-v^{2}}
{4c_{i}^{2}});\\ 
\frac{n}{n_{i}}=\exp(\frac{c_{i}^{2}-v^{2}}{2c_{i}^{2}})\ ,
\end{equation} \label{eq:a13}
where $v$ and $n$ are the velocity and the number density at the distance $r$ 
from the cloud centre. The outer limit of the cloud is given
by the location of the ionization front at $r=r_{i}$. 
Fig.~\ref{fig:globulenew} illustrates this model.

In the ionization front the equilibrium between ionization and recombination
is governed by the following equation:
\begin{equation}
\sigma_{i}J(1-x)=\alpha n x^{2}\ ,
\end{equation} \label{eq:a14}
where $x$ is the degree of hydrogen ionization, $n$ the hydrogen density,
$J$ the number of incident UV photons per cm$^{2}$ and s, $\alpha$
the recombination coefficient, and $\sigma_{i}$ 
the ionization cross section. The ionization cross section 
is calculated using the 
`on the spot' approximation. It is only valid if the recombination time
scale is much less than the kinematic timescale. This assumption
is justified in section 5.3. $J(r)$ is given by
\begin{equation}
J(r)=J_{0}e^{-(\tau_{g}(r)+\tau_{d}(r))}\ ,
\end{equation} \label{eq:a15}
where $\tau_{g}(r)$ and $\tau_{d}(r)$ are the optical depths of the 
gas and the dust along a radial line from an outer radius $R_{b}$
to an inner radius $r$. The optical depth of the gas is:
\begin{equation}
\tau_{g}(r)=\int_{r}^{R_{b}} n \sigma_{i}(1-x)dr=\int_{r}^{R_{b}}
\frac{\alpha n^{2}}{J}x^{2}dr\ .
\end{equation}  \label{eq:taug}
The optical depth of the dust is:
\begin{equation}
\tau_{d}(r)=\sigma_{d}\int_{r}^{R_{b}}ndr\ ,
\end{equation} \label{eq:a16}
where $\sigma_{d}$ is the UV cross section of the dust per hydrogen nucleus.
Differentiating equation \ref{eq:taug} 
with respect to $r$ and substituting $J(r)$ gives
\begin{equation}
J_{0}e^{-\tau_{g}(r)}d\tau_{g} = -\alpha n^{2}e^{\tau_{d}}dr\ .
\end{equation} \label{eq:a17}
Integrating this equation from $r_{i}$ to infinity and setting
$J(r_{i})=n_{i}c_{i}$ leads to
\begin{equation}
J_{0}=n_{i}c_{i}e^{\tau_{d}(r_{i})}+\alpha \int_{r_{i}}^{\infty}n^{2}
e^{\tau_{d}(r)}dr\ .
\end{equation} \label{eq:a18}
This is the final equation governing the radiation transfer in the ionization front.

Since the ionization front velocity is assumed to be small compared to the
local sound velocity of the neutral gas, the isothermal neutral 
condensation is in a quasi-equilibrium state. Its structure is
described by the Lane-Emden equation for isothermal spheres:
\begin{equation}
\frac{d^{2}u}{dx^{2}}+\frac{2}{x}\frac{du}{dx}=e^{-u}\ ,
\end{equation} \label{eq:a19}
where $x=\frac{r_{i}}{c_{n}}(4\pi G \rho_{c})^{\frac{1}{2}},$
$\rho_{c}$ is the cloud's central density and $G$ is the 
gravitation constant. The radial density is given by
\begin{equation}
\rho (x)=\rho_{c}e^{-u}\ .
\end{equation} \label{eq:a20}
The jump conditions over the D-critical ionization front lead to the following
expression for the gas pressure:
\begin{equation}
p_{n}(x)=\rho(x)c_{n}^{2}=\rho_{c}e^{-u}c_{n}^{2}=2p_{i}(x)=2\rho_{i}
c_{i}^{2}\ .
\end{equation} \label{eq:a21}
Since the density never drops to zero, an isolated isothermal sphere 
has no boundary. This changes with an external radiation field.
The outer boundary is given by the location of the ionization front at the
illuminated side and by the pressure of the surrounding ionized low density gas
at the shadowed side. It should be mentioned here that clouds with $x>6.5$ are 
gravitationally unstable (Jeans unstable).

\subsection{The disk model}

The disk is assumed to be continuous and smooth. We follow Pringle (1981) 
for the disk equations, replacing the $\alpha$-viscosity prescription
by the one described in section 3. Furthermore, the pressure due to the 
turbulent viscosity is added to the thermal pressure:
\begin{equation}
\nu=\frac{1}{Re}\xi^{\frac{1}{3}}H^{\frac{4}{3}}\rho^
{\frac{1}{3}}\ ,
\end{equation} \label{eq:a22}
\begin{equation}
\Sigma = Re\ \xi^{-\frac{1}{3}}H^{-\frac{4}{3}}\rho^{-\frac{1}{3}}
\frac{\dot{M}}{3\pi}\ ,
\end{equation} \label{eq:a23}
\begin{equation}
\rho=\Sigma/H\ ,
\end{equation} \label{eq:a24}
\begin{equation}
p=\rho G M(R) (\frac{1}{R} - \frac{1}{\sqrt{R^{2}+
\frac{H^{2}}{4}}})\ ,
\end{equation} \label{eq:a25}
\begin{equation}
p=\frac{\rho k_{\rm B} T_{\rm n}}{m_{p}} + \xi^{\frac{2}{3}}
\rho^{\frac{5}{3}} H^{\frac{2}{3}}\ .
\end{equation} \label{eq:a26}
The disk parameters are the height $H$, the surface density $\Sigma$, the mass accretion 
rate $\dot{M}$, and the neutral gas temperature $T_{\rm n}$; $k_{\rm B}$ is 
the Boltzmann constant.   
Here $\xi$ is defined by the equation $\epsilon=\xi\cdot\rho$.
The second term in equation (28) represents the turbulent pressure of
the clouds $p=\rho v_{\rm turb}^{2}$. The radial mass distribution
$M(R)$ is given explicitly by $M(R)=M+M_{0}R^{\frac{5}{4}}$, where
$M=3\, 10^{6}$ M$_{\odot}$ is the central mass component and
$M_{0}= 1.6\, 10^{6}\ {\rm M}_{\odot}/{\rm pc}^{\frac{5}{4}}$ 
describes the mass distribution of the stellar content. This is close to the findings
of Eckart \& Genzel (1996).
In addition we assume that the central density of the disk
in the continuous picture is linearly coupled to the central density
of the clouds in the discrete picture, i.e. $\rho=\beta \cdot \rho_{C}$.

Finally we take into account that the clouds are illuminated 
from the centre. Therefore the ionization front exists only on the
illuminated side. As described above, there is a flux of electrons
away from the clouds. These electrons fill up the space between
the clouds and thus build an intercloud medium which is the
Sgr A West H{\sc ii} region. Most of its mass is concentrated
within the disk plane. Its pressure
is responsible for the cloud boundary at the shadowed side. 
The electron density of the intercloud medium is calculated
using $H_{i}=c_{i}/\Omega$ and $\rho_{i}=\beta \Sigma/H_{i}$,
where $\Omega$ is the angular velocity of the clouds at a distance $R$
from the Galactic Centre.

\subsection{The physical conditions}

\begin{figure}
	\resizebox{\hsize}{!}{\includegraphics{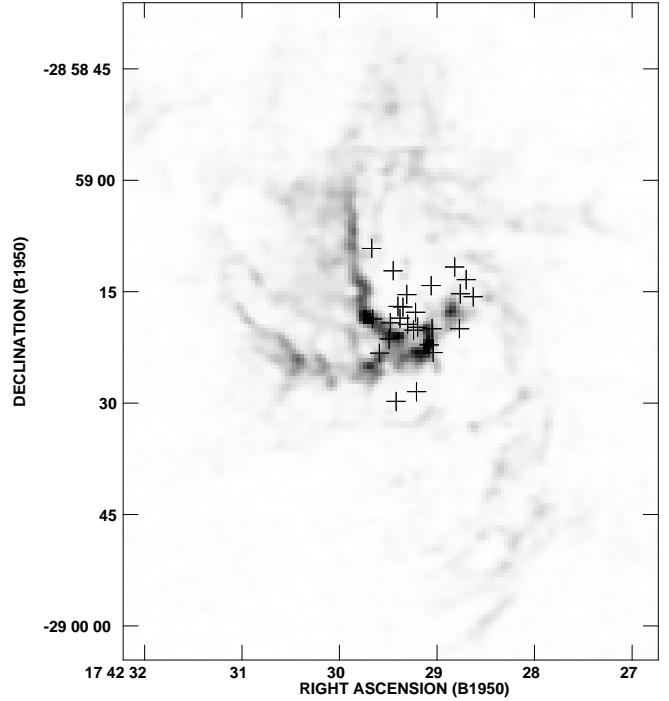}}
      \caption{Crosses: The He{\sc i} stars identified by
	Genzel et al. (1996). Greyscale: Ne{\sc ii} line emission 
	(Lacy et al. 1991).}
	\label{fig:stars}
\end{figure}

\begin{figure}
	\resizebox{\hsize}{!}{\includegraphics{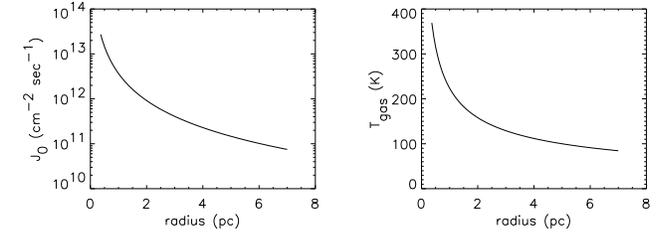}}
      \caption{The external radiation field and the inner gas temperature
	of the clouds as a function of the distance to the Galactic Centre.
} \label{fig:init}
\end{figure}

The radiation field in the Galactic Centre seems to be made
by the central cluster of He{\sc i} stars (Genzel et al. 1996).
In Fig.~\ref{fig:stars} we show the identified He{\sc i} stars together with
the Ne{\sc ii} data from Lacy at al. (1991). 
The clustering seen in Fig.~\ref{fig:stars} leads us to the assumption 
that the produced radiation field is approximately spherically symmetric
and follows a $R^{-2}$-law. The Lyman continuum (Lyc-)photon production rate
within 1 pc is $N_{\rm Lyc}\sim 10^{50}$\,s$^{-1}$ (Mezger, Duschl, \&
Zylka 1996).
Since it is assumed that this radiation is responsible for the 
heating of the clouds, the temperature follows a $R^{-\frac{1}{2}}$-law
(for direct heating see Puget \& Boulanger 1985). The absolute values of
the temperature are in agreement with molecular excitation models
(Sutton et al. 1990, Jackson et al. 1993). The FUV radiation field
is $G_{0}\sim 10^{5}$ (Wolfire et al. 1990) at 1 pc which is in
excellent agreement with the Lyc-photon production rate if one assumes
a black body temperature of 35\,000 K for this radiation.
The number of UV photons per cm$^{2}$ and s and the gas temperature
can be seen in Fig.~\ref{fig:init}.

The ambient H{\sc ii} region has a constant temperature of $T_{e}$=7000 K
(Roberts \& Goss 1993). The deduced isothermal sound speed in
the ionized medium is $c_{i}=7.6\, 10^{5}$ cm\,s$^{-1}$.
The recombination coefficient for hydrogen atoms is $\alpha=3.3\,
10^{-13}$ cm$^{3}$\,s$^{-1}$, the UV cross section per hydrogen nucleus
is $\sigma_{d}=1.11\, 10^{-21}$ cm$^{2}$. 
The Reynolds number is $Re=1000$, the critical value for the onset
of turbulence in laboratory experiments. For the factor $\beta$ between the
disk's central density and the clouds' central density the observed volume
filling factor is used: $\beta=\Phi_{\rm V}=\frac{1}{100}$.

\section{Results and verification of the assumptions} 
 
The problem which must be solved consists of a system of 
6 non-linear equations (equations (20), (24)-(28)) together with the 
Lane-Emden equation (21). It is solved using a Newton-Raphson iteration.
Given the non-linearity of the equations multiple solutions are possible.
We solved the equations with the parameters given above starting the
Newton-Raphson iteration from different initial values in order
to recover all solutions. These solutions represent a stationary
clumpy accretion disk which consists of distinct clouds.

It turns out that we can solve the
system of equations exactly and that there are two different
solutions for the radiation transfer with the clouds' central density 
almost unchanged. 
We refer to the two different solutions as the {\it heavy cloud} and 
the {\it light cloud} solutions. Since the only parameter which is not
a priori determined is the mass accretion rate, we varied this parameter
in the range 10$^{-6}$ M$_{\odot}$\,yr$^{-1} \leq \dot{\rm M} 
\leq10^{-1}$ M$_{\odot}$\,yr$^{-1}$. Higher values of 
$\dot{\rm M}$ give higher central densities and larger disk heights $H$. 
The adopted value of $\dot{\rm M}=10^{-4}$ M$_{\odot}$\,yr$^{-1}$ is a
compromise between a high $\dot{\rm M}$ leading to high central densities 
in order to resist the tidal shear and a low $\dot{\rm M}$ giving the 
observed small disk height.
Both solutions describe stable gas clouds in the sense discussed in section 2.

\subsection{The heavy clouds}
   
\begin{figure}
	\resizebox{\hsize}{!}{\includegraphics{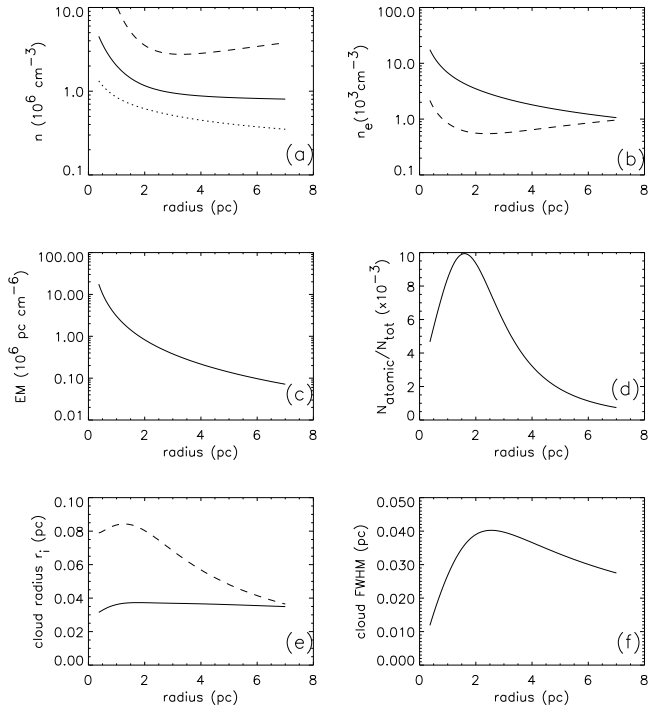}}
      \caption{The first set of results for the heavy clouds
	versus the distance to the Galactic Centre.
	(a) Solid: mean density; dotted: density at the cloud surface; dashed:
	central density. (b) Electron density at the outer radius.
	Solid: illuminated side; dashed: shadowed side. (c) Emission
	measure. (d) Atomic gas fraction. (e) Cloud radius. Solid:
	illuminated side; dashed: shadowed side. (f) Cloud
	diameter at half maximum density. 
} \label{fig:resulti}
\end{figure}

\begin{figure}
	\resizebox{\hsize}{!}{\includegraphics{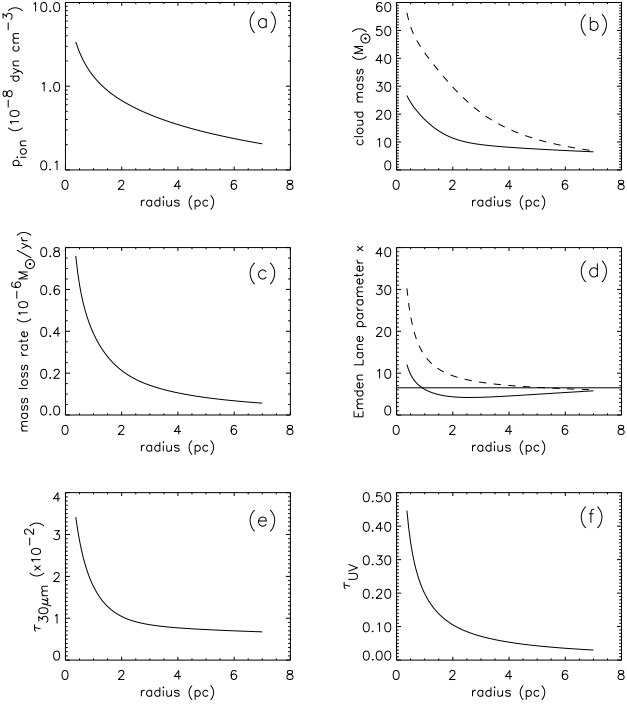}}
      \caption{The second set of results for the heavy clouds
	versus the distance to the Galactic Centre.
	(a) Pressure of the ionized gas at the outer rim at the
	illuminated side. (b) Total mass. Solid: calculated with
	the radius of the illuminated side. Dashed: calculated
	with the radius of the shadowed side. (c) Mass loss rate
	due to the ionization front. (d) Emden Lane parameter.
	Solid:  calculated with the radius of the illuminated side. 
	Dashed: calculated with the radius of the shadowed side.
	Horizontal line: limit for gravitational collapse x=6.5.
	(e) Optical depth at 30$\mu$m. (f) UV optical depth. 
} \label{fig:resultii}
\end{figure}

Fig.~\ref{fig:resulti} and \ref{fig:resultii} show the results 
for the heavy clouds.
Plot \ref{fig:resulti}(a) shows the number densities of the neutral gas 
inside the cloud.
The averaged density is plotted
with a solid line, the central density with a dashed line, and the 
density at the cloud surface with a dotted line. It is remarkable that the central
density has an almost constant value for distances greater than 
3 pc and that there is more than a factor of
6 between the surface density and the central density. At a distance 
to the Galactic Centre of 2 pc the central density is about $\sim3.5\,
10^{6}$ cm$^{-3}$, corresponding very well to the value
derived from HCN excitation models (Jackson et al. 1993), 
whereas the surface density has a value of $\sim 6\, 10^{5}$ cm$^{-3}$.
Since the gas in the PDR is hotter and therefore the gas density is 
expected to be lower than derived in this simple isothermal model,
the value stated above corresponds well to the value derived from the
[O{\sc i}] 63 $\mu$m/[C{\sc ii}] 158 $\mu$m ratio and the H$_{2}$
1-0/2-1 S(1) ratio (Burton et al. 1990, Wolfire et al. 1990)
and from the low-J CO line emission (see e.g. Sutton et al. 1990).   
Thus, the HCN line emission
originates in the dense cloud core, whereas the low-J CO and the H$_{2}$ S(1) lines 
are emitted in the atomic/molecular transition region of the PDR.

Plot \ref{fig:resulti}(b) shows the electron density at the cloud surface as a 
solid line and the electron density of the intercloud medium as
a dashed line. The latter is in good agreement with the value
derived from $\lambda$19 and  [S{\sc iii}] 3 $\mu$m line ratios
(Erickson et al. 1994). The electron density at the surface lies exactly
in the range given by Lo \& Claussen (1983). We conclude that
the radio continuum and line emission of high surface brightness 
at the inner edge of the CND originates in the cloud surfaces and the low brightness
overall emission distribution corresponds to the low density intercloud medium. 
Both features form the H{\sc ii} region Sgr A West.

Plot \ref{fig:resulti}(c) shows the emission measure 
$EM=\int_{r_{i}}^{+\infty} n_{i}^{2} dr$. It is in agreement with the value
given by Beckert et al. (1996) based on the 15 GHz flux
of Brown \& Liszt (1984). The ratio between the column density of the atomic gas
and the total column density $\eta=
N_{\rm atomic} / \overline{N_{\rm tot}}$ (plot \ref{fig:resulti}d) is determined 
by the dust absorption
for distances smaller than 2 pc and by H$_{2}$ self-shielding for
larger distances. Estimating the atomic hydrogen column density
at 2 pc $N_{\rm atomic} \simeq 2\eta \overline{n}r_{i}$ gives a value of
$\sim 4\, 10^{21}$ cm$^{-2}$. The observed value (Liszt et al. 1983)
is about twice this estimate.  
We observe a significant difference of the cloud radius \ref{fig:resulti}(e)
between the illuminated and the shadowed side. 
We can interpret them as upper and lower limits for the cloud radius.

The two graphs in plot \ref{fig:resultii}(b) can be interpreted 
as upper and lower limits for the cloud mass, 
which lies in a range between 10 and 30
M$_{\odot}$. The timescale for the mass loss (plot \ref{fig:resultii}c) 
by the outward electron flux is about 10$^{8}$ yr for a 
cloud mass of 10 M$_{\odot}$. The clouds
are thus in a quasi stationary state. An important plot is the 
one showing the Lane-Emden parameter (plot \ref{fig:resultii}d). 
The dashed line corresponds to
the shadowed side, the solid line to the illuminated side.
Interpreting these two lines again as upper and lower limits for the actual
$x$, we find a tendency that the clouds become gravitationally unstable
for distances to the Galactic Centre of less than about 2 pc. 
In order to compare our findings with observations, we also plot the
30 $\mu$m optical depth (plot \ref{fig:resultii}e)
$\tau_{30\mu m}\simeq 2\overline{n}r_{i}\kappa_{30\mu m}$ where the 
absorption coefficient $\kappa_{30\mu m}$ is taken from Draine (1985). 
This can be directly compared with the 30$\mu$m measurements of 
Telesco et al. (1996), who measured an optical depth at 30$\mu$m
of 10$^{-2}$ at the outer edge of the minispiral. This compares very well 
with the outcome of our model (Fig.~\ref{fig:resultii}e).
The UV optical depth across the ionization front $\tau_{\rm UV}$ 
(plot \ref{fig:resultii}f) tells us that the
ionization front boundary is determined by atomic hydrogen self-shielding.

\begin{figure}
	\resizebox{\hsize}{!}{\includegraphics{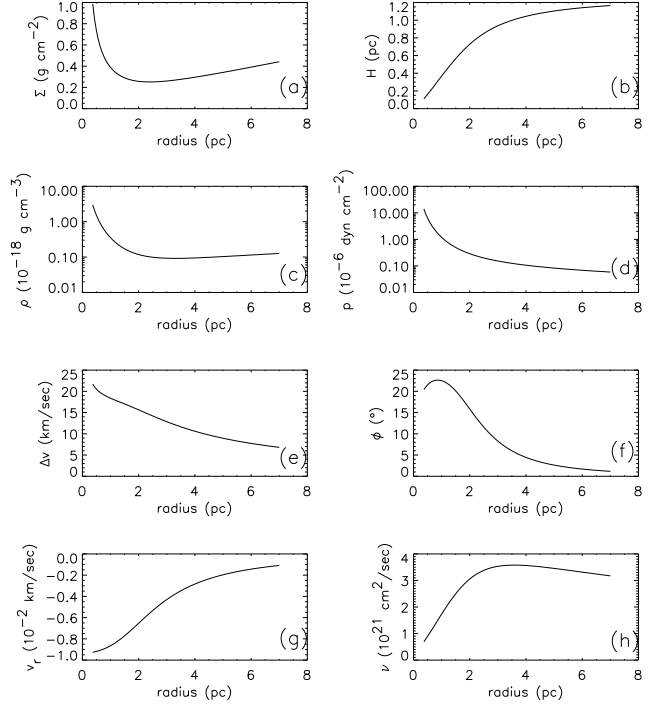}}
      \caption{\label{fig:scheibei} The disk parameters versus the distance to the Galactic Centre.
	(a) Total gas surface density.
	(b) Disk height. (c) Central gas density. (d) Gas pressure.
	(e) Dispersion velocity of the clouds. (f) Disk flaring angle.
	(g) Accretion velocity. (h) Viscosity.
} 
\end{figure}

We now examine the solution for the averaged parameters of the disk 
model (Fig.~\ref{fig:scheibei}). 
The parameters shown are the column density $\Sigma$ 
(plot \ref{fig:scheibei}a),
the disk height H (plot \ref{fig:scheibei}b), the disk's central density 
(plot \ref{fig:scheibei}c), the central pressure (plot \ref{fig:scheibei}d),
the velocity dispersion $\Delta v$ of the clouds (plot \ref{fig:scheibei}e), 
the disk flaring angle $\Phi$ (plot \ref{fig:scheibei}f), the radial inward 
velocity $v_{r}$ (plot \ref{fig:scheibei}g), and the turbulent
viscosity $\nu$ (plot \ref{fig:scheibei}h). Since these clouds are only 
found at distances
larger than 2 pc, we will only comment on this part of the graphs.
It is remarkable that the disk height fits well with the one deduced
from observations (G\"{u}sten et al. 1987). The central density, which
is assumed to be proportional to the central density of a cloud, stays
approximately constant, which is a reasonable result. 
The velocity dispersion of the clouds
lies in a range between 10 and 20 km\,s$^{-1}$, compared to 30 km\,s$^{-1}$ 
derived by the same authors. The accretion velocity $v_{r}$ is
extremely low, which means that the clouds take a long time to approach
the centre. 

The viscosity $\nu$ used here is equivalent to an effective $\alpha$
as used in the standard prescription $\nu=\alpha c_{\rm n} H$.
It has a nearly constant value of $\alpha\sim 0.01$.

The cloud inflow time is given by Krolik \& Begelman (1988) as
\begin{equation}
t_{\rm inflow}\sim (\Delta v/v_{\rm orb})^{-2} t_{\rm coll}\ ,
\end{equation} 
where we take the cloud collision rate derived at the end of this section.
this leads to a mass accretion rate 
of $\dot{M}\sim M_{\rm tot}/t_{\rm inflow}$. Assuming a total disk mass
of $M_{\rm tot}=5\, 10^{4}$M$_{\odot}$ gives $\dot{M}\sim 10^{-4}$
M$_{\odot}$\,yr$^{-1}$. 
This value is in good agreement with the mass accretion rate used 
in our model.

\subsection{The light clouds}

\begin{figure}
	\resizebox{\hsize}{!}{\includegraphics{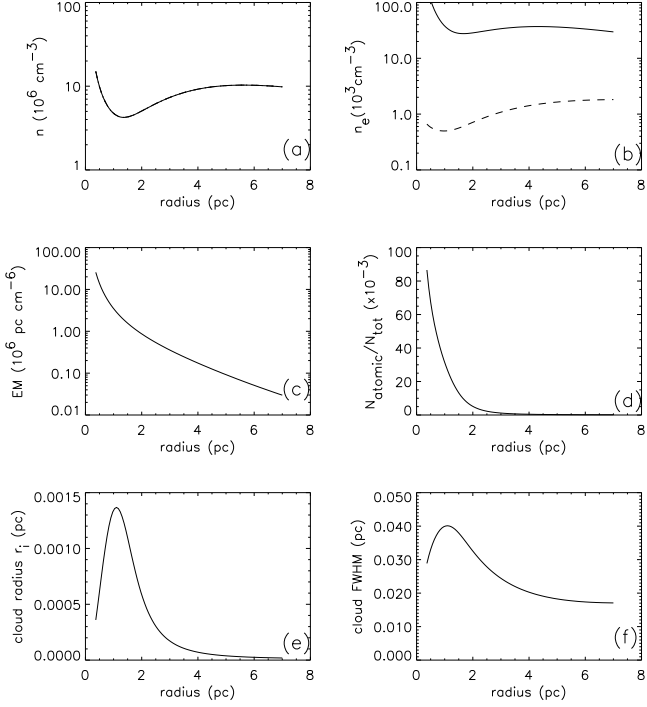}}
      \caption{The first set of results for the light clouds
	versus the distance to the Galactic Centre.
	 (a) Mean density (b) Electron density at the outer radius.
	Solid: illuminated side; dashed: shadowed side. (c) Emission
	measure. (d) Atomic gas fraction. (e) Outer radius at the
	illuminated side (f) Cloud diameter at half maximum density.
} \label{fig:leichtei}
\end{figure}

\begin{figure}
	\resizebox{\hsize}{!}{\includegraphics{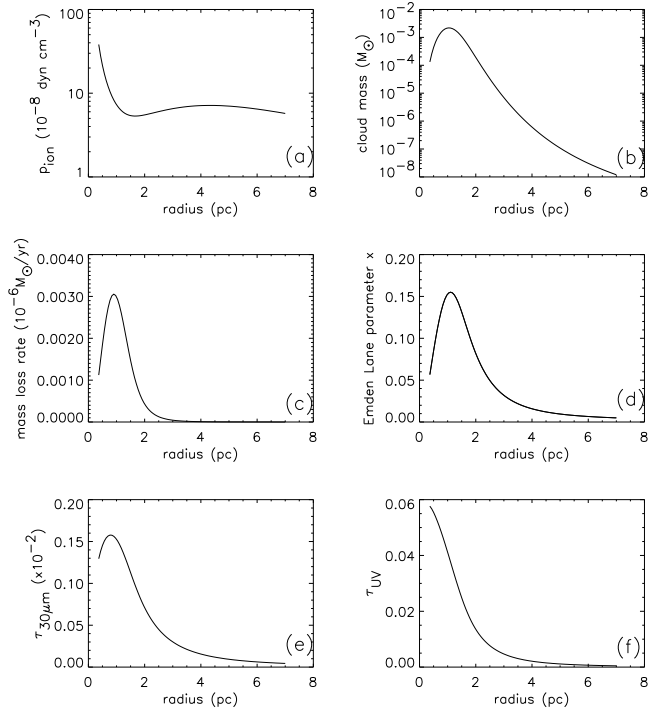}}
      \caption{The second set of results for the light clouds
	versus the distance to the Galactic Centre.
	(a) Pressure of the ionized gas at the outer rim of the
	illuminated side. (b) Total mass calculated with
	the radius of the illuminated side. (c) Mass loss rate
	due to the ionization front. (d) Emden Lane parameter
	calculated with the radius of the illuminated side.
	(e) Optical depth at 30$\mu$m. (f) UV optical depth.
} \label{fig:leichteii}
\end{figure}

Figures \ref{fig:leichtei} and \ref{fig:leichteii} show the same 
parameters as Figures \ref{fig:resulti} and 
\ref{fig:resultii} but for another class of solutions.
Here the densities of the cloud centre (plot \ref{fig:leichtei}a),
at the outer boundary and the mean density have the same value than that 
of the heavy clouds.
The observed increase of the density with increasing distance
might not be significant, whereas its increase with 
decreasing distance smaller than 1 pc is important because 
only denser clouds
can resist tidal shear at smaller distances.  
The clouds are  100 times smaller (plot \ref{fig:leichtei}e) and 1000 
times lighter (plot \ref{fig:leichteii}b) than
the heavy ones. 
Thus, the light clouds have the appearance of stripped cores of the heavy clouds. 

Their distribution shows a maximum of  
radius and mass at a distance of about 1 pc. The column density
of atomic hydrogen (plot \ref{fig:leichtei}d) is determined everywhere by 
H$_{2}$/CO self-shielding
because of the high density at the outer boundary. The electron density
at the illuminated boundary is 10 times higher than that of the heavy clouds
(plot \ref{fig:leichtei}b), whereas the emission measure is comparable to 
that of the heavy clouds (plot \ref{fig:leichtei}c). 
They evaporate much faster ($t_{\rm evap}\sim 10^{6}$ 
yr at 2 pc) (plot \ref{fig:leichteii}c)
and are gravitationally stable ($x < 6.5$) (plot \ref{fig:leichteii}d).

These clouds might account for the high-J CO lines which require
a density close to 10$^{7}$~cm$^{-3}$ in the atomic/molecular transition region
(Burton et al. 1990).

\subsection{Verification of the assumptions}

In this section we show that our model assumptions are valid for
the light and heavy clouds.
We only discuss the results for the heavy clouds because they contain most
of the mass of the CND. 

\begin{figure}
	\resizebox{\hsize}{!}{\includegraphics{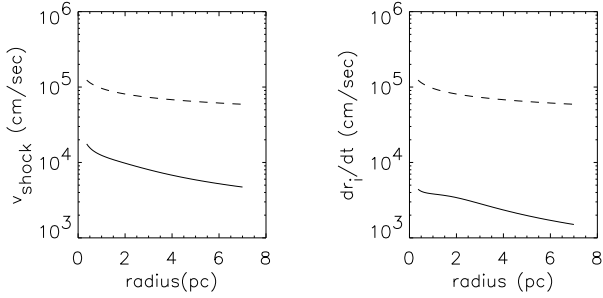}}
      \caption{Heavy clouds properties versus the distance to 
	the Galactic Centre:
	left: the sound velocity in the neutral gas (dashed line)
	together with the velocity of the shock front $v_{\rm shock}$
	(solid line). 
	Right: the sound velocity in the neutral gas (dashed line)
	together with the velocity of the ionization front $v_{\rm IF}$ 
	(solid line).
} \label{fig:shock}
\end{figure}

As described above, the outer layer of the cloud has three different
regions. The ionization front, caused by UV photons with energies
above 13.6 eV, is located outside, followed by the PDR whose origin is the 
FUV radiation
field, followed by a shock front due to the higher temperature and therefore
higher pressure in the PDR. The cloud can only be assumed to be stationary
if the velocity of the shock front $v_{\rm shock}$ and the velocity of
the ionization front $v_{\rm IF}$ are much smaller than the sound velocity within 
the cloud. In this case they do not affect the clouds' overall structure.
(Dyson 1968). The velocity of the shock front is given by 
(Bertoldi \& Draine 1996)
\begin{equation}
v_{\rm shock}\simeq c_{n}(99\delta )^{\frac{1}{4}}e^{-\frac
{\tau_{Ly}}{4}}(1+\frac{\tau_{Ly}}{3})^{\frac{1}{4}}(1+0.72\frac{r_{i}}{R})^
{\frac{1}{4}}\ ,
\end{equation} \label{eq:a29}
where $\delta=J_{0}/(2\alpha n_{n} r_{i})$ is the `Str\"{o}mgren number',
R the distance to the Galactic Centre, and $\tau_{Ly}$ is the UV optical depth.
The velocity of the ionization front can be obtained by $v_{\rm IF}\sim r_{i}\dot{M}/M$
where $M$ is the cloud mass and $\dot{M} \simeq 4\pi r_{i}^{2}c_{i}n_{i}$.  
Fig.~\ref{fig:shock} shows these velocities
together with the sound velocity in the neutral gas.
The sound velocity is always a factor 10 larger than that of the shock 
front and the ionization front velocity. Thus, the assumption of a quasi-stationary 
equilibrium is justified. In addition, it should be noticed that the
velocity of the shock front is larger than that of the ionization front.

The assumption of an isothermal ionization front only holds if the characteristic 
cooling length is small against the depth of the ionized layer.
The characteristic cooling time in the H{\sc ii} region outside the cloud is 
$t_{\rm cool}=6.3\, 10^{11}/n_{i}$ s. With a typical electron
density of $\sim 10^{4}$ cm$^{-3}$ we obtain a typical cooling length
of $L_{\rm cool}=c_{i}t_{\rm cool}=4.8\, 10^{13}$ cm. This is
$\sim 10^{-4}$ times the cloud radius. If the ionization front has a 
depth of A$_{\rm V}\sim0.01$ (Tielens \& Hollenbach 1985), its extent
is $\sim 10^{-3}$ times the cloud radius.

The assumption of an ionization-recombination equilibrium is only
valid if the timescale of the recombination is small against the
kinematic timescale. The recombination timescale at 7000 K is given
by Spitzer (1978): $t_{\rm rad}=3\, 10^{12}/n_{\rm e}$ s.
With the same electron density used above one obtains $t_{\rm rad}=3\,
10^{8}$ s. The kinematic timescale is approximately given by 
$t_{\rm kin}\sim r_{i}/c_{i}$ which gives a value of 
$t_{\rm kin}\sim 10^{11}$ s.

\begin{figure}
	\resizebox{\hsize}{!}{\includegraphics{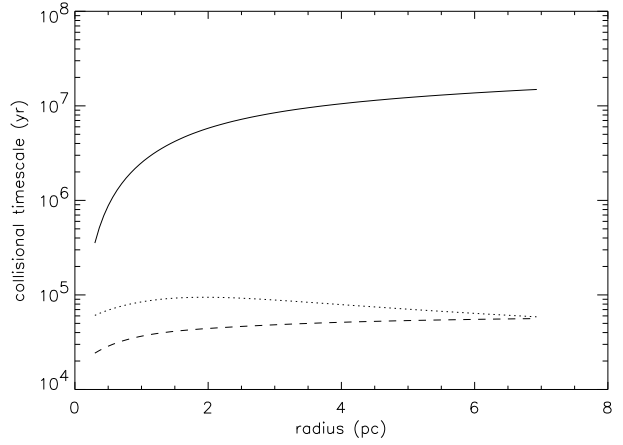}}
      \caption{Timescales for the heavy clouds versus the distance 
	to the Galactic Centre:
	Solid line: Collisional timescale due to cloud cloud 
	collisions. Dashed line: Thermal timescale using the cloud 
	radius of the
	illuminated side. Dotted line: Thermal timescale using the cloud 
	radius of the shadowed side. 
} \label{fig:collzeit}
\end{figure}

Furthermore, we must check if the collisional timescale is much smaller
than the crossing time for a sound wave (thermal timescale). 
Only in this case does the cloud have enough time to reach the 
quasi-equilibrium state. The thermal timescale is given by $t_{\rm thermal}=
r_{i}/c_{\rm n}$. For the collisional timescale
we choose the expression $\nu \sim v_{\rm turb}^{2}\big(t_{\rm coll} 
\Omega^{2}\big)^{-1}$ (Pringle 1981). In this case it is assumed that the 
collisional timescale is larger than the orbital timescale
(Shlosman et al. 1990) and the disk can be treated in analogy
to the Saturn's rings (Goldreich \& Tremaine 1978).
As the CND has a volume filling factor of 0.01 we think that this 
assumption is justified. 
Nevertheless, it has to be proven by dynamical models, which is 
beyond the scope of this paper. Such a numerical model,
which will be discussed in a following paper, seems to agree 
with this assumption.
The different timescales are shown in Fig.~\ref{fig:collzeit}.  
Clearly, the collisional timescale is much larger than the thermal one.
It is remarkable how large this timescale is. In this scenario
there is only one collision every 10$^{4}$ yr in the whole isolated CND.

We made the assumption that the time and space averaged dissipation 
rate in the hot, layers during a cloud--cloud collision can be approximated 
by the C{\sc ii} line emission of the smoothed large scale disk.
In the hot partially ionized layer which is formed during a cloud--cloud
collision the gas has a density $\geq 10^{5}$ cm$^{-3}$ and is heated to 
a temperature of $\sim$1000 K. Under these conditions the O{\sc i}, C{\sc ii},
and H$_{2}$ lines are mainly responsible for the cooling of the gas. 
The total cooling rate in this layer
is approximately $\Lambda \sim 10^{-17}$ erg\,cm$^{-3}$\,s$^{-1}$
(Tielens \& Hollenbach 1985). The C{\sc ii} cooling rate of our 
smooth model disk is 
$\Lambda_{\rm CII} = \epsilon\,\rho_{\rm C} \sim 10^{-19}$ 
erg\,cm$^{-3}$\,s$^{-1}$. 
We now have to account for the transient character of the hot layer.
Its lifetime can be approximated by the thermal timescale $t_{\rm thermal}$
of the cloud
compared to the timescale between two collisions. The transient and continuous
cooling rates are thus connected in the following way:
\begin{equation}
\Lambda \sim \Lambda_{\rm CII}\,\frac{t_{\rm coll}}{t_{\rm thermal}}\ .
\end{equation}
Fig.~\ref{fig:collzeit} shows that $t_{\rm coll}/t_{\rm thermal} \sim
100$. This justifys our assumption.

\section{Tidal forces}

\begin{figure}
	\resizebox{\hsize}{!}{\includegraphics{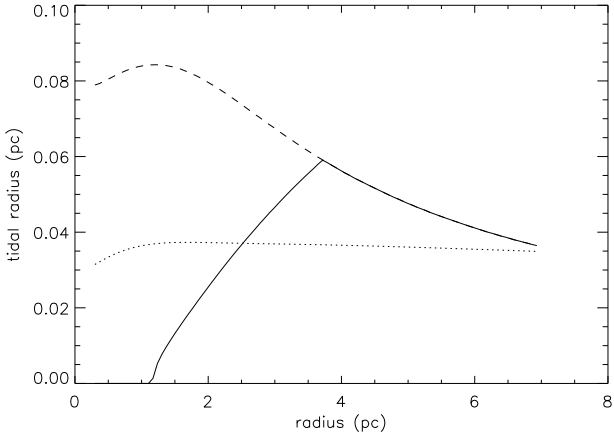}}
      \caption{Dimensions of the heavy clouds versus the distance 
	to the Galactic Centre. Solid line: the tidal radius $r_{\rm cl}$.
	Dashed line: the cloud radius at the shadowed side. 
	Dotted line: the cloud radius at the illuminated side.
} \label{fig:gezeiten}
\end{figure}

The clouds in the vicinity of the Galactic Centre are exposed to
strong tidal forces which tend to disrupt the clouds, 
except if self-gravity is effective against tidal shear.
The tidal (Roche) limit for the cloud mass $M_{\rm cl}$ is given by 
(Mathews \& Murray 1987)
\begin{equation}
\frac{3}{5} \frac{G M_{\rm cl}^{2}}{r_{\rm cl}} > \frac{1}{5} M_{\rm cl} r_{\rm cl}^{2}|f'(R)|\ ,
\end{equation} \label{eq:a30}
where $f(R)=GM(R)/R^{2}$, $r_{\rm cl}$ the cloud radius, and $R$ is the cloud's
distance to the Galactic Centre.
We have calculated the cloud tidal radius $r_{\rm cl}$
analytically. For the heavy clouds it is compared to the cloud radii at the 
illuminated and the shadowed side (Fig.~\ref{fig:gezeiten}). 
The tidal radius is smaller than the radius determined 
by the UV radiation field at distances less than 2.5\,pc and smaller
than the radius of the shadowed side at distances less than 3.5\,pc.

\begin{figure}
	\resizebox{\hsize}{!}{\includegraphics{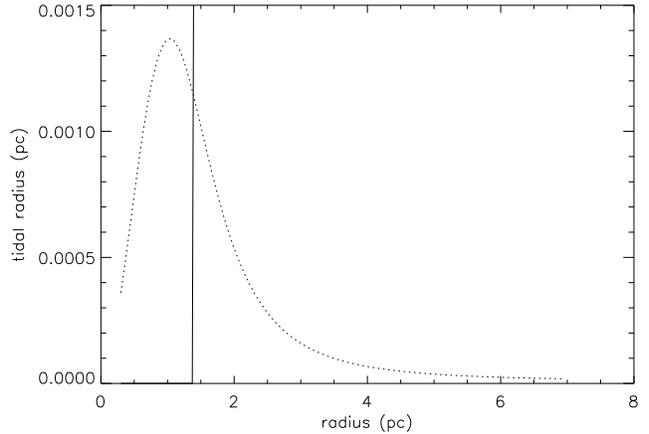}}
      \caption{Dimensions of the light clouds versus the distance 
	to the Galactic Centre.
	Solid line: the tidal radius $r_{\rm cl}$
	of the light clouds.
	Dotted line: the cloud radius at the illuminated side. 
} \label{fig:gezeitenleicht}
\end{figure}

We have also calculated the cloud tidal radius $r_{\rm cl}$ analytically 
for the light clouds. It is compared to the cloud radii at the 
illuminated and the shadowed side in Fig.~\ref{fig:gezeitenleicht}.
These very small clouds are stable against tidal shear for distances
from the Galactic Centre greater than 1.5\,pc.
Comparing the gravitational collapse time $t_{\rm ff}$ to the orbital 
period $P_{\rm orb}$ at a distance of $R_{\rm Centre}$=1\,pc from the Galactic Centre yields:
\begin{equation}
\frac{t_{\rm ff}}{P_{\rm orb}} \simeq 0.5\Big(\frac{n}{10^{7}\ {\rm cm}^{-3}}\Big)^
{-\frac{1}{2}}R_{\rm centre}^{-1}\Big(\frac{v_{\rm orb}}{150\ {\rm km\,s}^{-1}}\Big)\ ,
\end{equation}
where $n$ is the gas density of the cloud and $v_{\rm orb}$ is the orbital velocity of the cloud.
Thus, these self-gravitating light clouds will be stretched out along their
orbital path by an amount comparable to their initial radial extent 
within about one orbital period.

\section{The mass accretion rate}

G\"{u}sten et al. (1987) have derived a mass accretion rate of the clouds
in the CND of $\dot{M} \sim 10^{-2}$\,M$_{\odot}$\,yr$^{-1}$. They assumed 
that the total luminosity in the infrared and submm-lines comes entirely 
from the dissipation of the turbulent motions. 

We suggest that if the turbulent energy dissipation is due to the infrared line emission
(for which we have used the C{\sc ii} line as representative), 
it should not exceed several percent of the whole
infrared line emission due to the PDRs. The total luminosity in the
infrared and submm-lines (PDR) is L=2$\,10^{4}$L$_{\odot}$ (Genzel 1989).
The emitted luminosity of the C{\sc ii}-line $L\sim \epsilon\times
M_{\rm tot}$ in our model represents $\sim$10\% of the total luminosity. 
This is consistent with the radiative dissipation rate given by Genzel et al. (1985)
\begin{equation}
L_{\rm shock} \sim 2\,10^{3}\big(\frac{\eta_{\rm s}}{0.1}\big)\big(\frac{M_{\rm CND}}
{10^{4}\,{\rm M}_{\odot}}\big)\big(\frac{\Delta v}{30\,{\rm km\,s}^{-1}}\big)^{2}\big(\frac{10^{4}\,{\rm yr}}{\tau_{\rm coll}}\big)\ {\rm L}_{\odot}\ ,
\end{equation}
where $\eta_{\rm s}$ is the fraction of kinetic energy converted into radiation
via shocks, $M_{\rm CND}$ is the mass of the CND, and $\Delta v$ is its velocity 
dispersion. We stress here that $\tau_{\rm coll}$ is the mean collision time
for all clouds in the CND in order to recover the total luminosity
due to radiative energy dissipation.

\section{Conclusions}

We have constructed an analytic model for the CND. It mainly
consists of $\sim$ 500 heavy clouds moving around the Galactic
Centre building a disk-like structure. The spatial structure of
the cloud ensemble is described by a smooth continuous disk whose
viscosity is due to partially inelastic cloud--cloud collisions.
This work represents a first attempt to model a clumped disk
analytically including a simple UV radiation transfer. The input
parameters ($T, J_{0}, \beta, Re$) are set to observed quantities.
At the chosen Reynolds number $Re$ turbulence sets in laboratory
experiments. The real unknown and therefore free parameter is the
mass accretion rate $\dot{\rm M}$ of the disk. The independent
resulting parameters ($n, n_{e}, EM, r_{i}, H, \Delta v$) are in
excellent agreement with observations at multiple wavelengths.

There are two solutions for our set of equations that correspond to
two stable clump regimes:
\begin{enumerate}
\item
the observed heavy molecular clouds with masses of $\sim$10
M$_{\odot}$ and sizes of 0.1\,pc,
\item
the stripped cores of the heavy clouds with masses between
10$^{-5}$ to 10$^{-4}$\,M$_{\odot}$ and sizes of $\sim
10^{-3}$\,pc.
\end{enumerate}

It will be of importance to find out whether---in our Galactic
Center and/or the centers of other galaxies---the light cloud 
population plays a role.

Within the disk, the number of collisions between clouds is very
low ($\overline{n}_{\rm coll} \sim 10^{-4}$ yr$^{-1}$ for about
500 clouds). We infer a mass accretion rate for the isolated CND
of $\dot{M} \simeq 10^{-4}$\,M$_{\odot}$\,yr$^{-1}$. We conclude
that the CND is much more stable and has a much longer lifetime
($\sim 10^{7}$ yr) than previously assumed.

In the present discussion we have focussed on the investigation 
of the two stable regimes of clouds (heavy and light). For a more
complete model of the CND, however, several aspects need to be
adressed additionally, including:
\begin{itemize}
\item The influence of tidal forces and internal rotation of the
clouds on their stability. While tidal forces tend to disrupt
the clouds, the role of rotation is not so clear, since rotation may
easily lead to modifications of the cloud structure and thus to
clouds that are more or less stable than the non-rotating ones.
The interplay between gravity, tidal forces, rotation, and the
presence of a point-like central mass distribution (i.e., the
central black hole) may explain why the CND seems to be almost 
empty of material in its inner $\approx 2$~pc.
\item The connection between the structure and dynamics of the gaseous 
material in the CND and the stellar component in the immediate
vicinity of the central black hole.
\end{itemize}

\begin{acknowledgements}
This work was supported by the {\it Deutsche Forschungsgemeinschaft 
(DFG)\/} through {\it Sonderforschungsbereich (SFB) 328 ``Evolution of
Galaxies''\/} at the University of Heidelberg. The authors thank an 
anonymous referee for helping to improve the presentation of the results
presented in this paper.
\end{acknowledgements}

\end{document}